\documentstyle[12pt]{article}

\def\bseq{\begin{subequation}}  % = 1a 1b
\def\eseq{\end{subequation}}
\def\bsea{\begin{subeqnarray}}  % = 1.1a 1.1b
\def\esea{\end{subeqnarray}}
\def\bbar{b \kern-.40em /}

\newcommand{\bbox}{\lower.2ex\hbox{$\Box$}}

\newcommand{\beq}{\begin{equation}}
\newcommand{\eeq}{\end{equation}}
\newcommand{\bea}{\begin{eqnarray}}
\newcommand{\eea}{\end{eqnarray}}
\newcommand{\ena}{\end{eqnarray}}

\newcommand{\wt}{\tilde}
\renewcommand{\a}{\alpha}
\renewcommand{\b}{\beta}

\newcommand{\th}{\theta}

\renewcommand{\l}{\left}

\renewcommand{\t}{\tau}

\newcommand{\r}{\right}
\newcommand{\nn}{\nonumber}
\def\II{\relax{\rm I\kern-.60em 1}}

\newcommand{\ba}{\begin{array}}
\newcommand{\ea}{\end{array}}

\begin{document}

\begin{titlepage}
\begin{flushright} IFUM--647--FT\\
%hep-th/99yyxxx\\
 \end{flushright}
\vfill
\begin{center}
{\LARGE\bf A $Z_{2} \times Z_{2}$ orientifold with spontaneously\\
\vskip 3.mm
broken supersymmetry}\\
\vskip 27.mm  
{\large\bf  A. L. Cotrone } \\
\vskip 5.mm
{\small
Dipartimento di Fisica dell'Universit\`a di Milano,\\
via Celoria 16,
I-20133 Milano, Italy}\\
\end{center}
\vfill

\begin{center}
{\bf ABSTRACT}
\end{center}
\begin{quote}
We present a $Z_{2} \times Z_{2}$ four dimensional orientifold in which 
supersymmetry is broken by a temperature-like Scherk-Schwarz mechanism.
As usual in this type of models, at the tree-level the breaking affects only the branes tangent to the direction involved by the SS deformation, and it can propagate only through radiative corrections to other sectors, where the gauge group is broken.
The result is a non-chiral model with gauge group $Usp(16)^{\otimes 2}\times Usp(8)^{\otimes 4}$.

\vfill     
\vskip 5.mm
 \hrule width 5.cm
\vskip 2.mm
{\small
\noindent e-mail: cotrone@pcteolaur1.mi.infn.it}
\end{quote}
\begin{flushleft}
Sept 1999
\end{flushleft}
\end{titlepage}

In the last year some authors \cite{sagnotti1}-\cite{sagnotti4} have shown how to build open string theories in which supersymmetry is totally or partially broken by the string anologous of the Scherk-Schwarz mechanism.
These models are of particular interest in relation with the various scenarios of compactification of string theories with large compact dimensions, which have possible fenomenological interest also at the energy of a few TeV \cite{grandi}\cite{antoniadis1}. 
In this context the Type I has proved to be a very flexible theory, where supersymmetry can be broken on some sets of D-branes, namely the ones that are tangent to the SS-breaking direction, while on the D-branes transverse to the SS direction it is unbroken, at least at the tree level.
On this last type of branes the breaking can be transmitted only by radiative corrections.
This general setting can accomodate scenarios where the scale of supersymmetry breaking, given by the size of the compactified dimension, demands for a string scale close to the electroweak energy in order to achieve realistic models; in this case the world we live in should be localized on the branes tangent to the SS direction.
If otherwise our world is supposed to be on the other type of branes, the string scale can be raised to intermediate values, for supersymmetry breaking would be mediated to us only by radiative corrections.
In this paper we build a model that provides an additional example of these phenomena, namely a Type I theory with three groups of D5-branes; two of these are ortogonal to the SS direction and have tree-level supersymmetry preserved, while supersymmetry is broken on the third group and on the D9-branes.\\ 
We start from the four dimensional compactification on \mbox{$T^{2}\times T^{2}\times T^{2}$} of the Type-IIB theory, orbifolded by the \mbox{$Z_{2}\times Z_{2}$} action generated by the identity (we will call it ``$o$'') and the $\pi$ rotations $g:(+,-,-)$, $f:(-,+,-)$, $h:(-,-,+)$, where the three entries in the parentheses refer to the three internal tori, while ``$+$'' and ``$-$'' are the two group elements of $Z_{2}$.
The torus partition function reads:
\bea \label{toron}
T &=&\frac{1}{4\t_{2}|\eta|^{4}}\ \Lambda_{1}[ \ba{c} h_{1} \\ g_{1} \ea ]\ \Lambda_{2}[ \ba{c} h_{2} \\ g_{2} \ea ]\ \Lambda_{3}[ \ba{c} h_{1}+h_{2} \\ g_{1}+g_{2} \ea ]\times \sum_{h_{1},g_{1},h_{2},g_{2}=0}^{\frac{1}{2}} \times\nn\\
&& \sum_{\a,\b=0}^{\frac{1}{2}}(-1)^{2(\a+\b+\a \b)}\  \frac{\th[\ba{c} \a \\ \b \ea] \ \th[\ba{c} \a+h_{1} \\ \b+g_{1} \ea]\ \th[\ba{c} \a+h_{2} \\ \b+g_{2} \ea]\ \th[\ba{c} \a-h_{1}-h_{2} \\ \b-g_{1}-g_{2} \ea]}{2\eta^{4}} \times \nonumber\\
&& \sum_{\wt{\a},\wt{\b}=0}^{\frac{1}{2}}(-1)^{2(\wt{\a}+\wt{\b}+\wt{\a}\wt{\b})}\  \frac{\wt{\th}[\ba{c} \wt{\a} \\ \wt{\b} \ea]\  \wt{\th}[\ba{c} \wt{\a}+h_{1} \\ \wt{\b}+g_{1} \ea]\ \wt{\th}[\ba{c} \wt{\a}+h_{2} \\ \wt{\b}+g_{2} \ea]\ \wt{\th}[\ba{c} \wt{\a}-h_{1}-h_{2} \\ \wt{\b}-g_{1}-g_{2} \ea]}{2\wt{\eta}^{4}}~,
\eea
where
\bea \label{bostwis}
\Lambda_{i}[ \ba{c} h \\ g \ea] &=& \ \Lambda_{i} \qquad if\ h=g=0 \\
\label{bostwist} \Lambda_{i}[ \ba{c} h \\ g \ea]&=&\ \frac{4|\eta|^{2}}{\th[ \ba{c} \frac{1}{2}-h \\ \frac{1}{2}-g \ea]\ \wt{\th}[\ba{c} \frac{1}{2}-h \\ \frac{1}{2}-g \ea]} \qquad otherwise
\eea
and $\Lambda_{i}$ are the lattice sums associated with the three tori.
The open descendant of this theory has been studied in \cite{berkooz1}; the open descendants of the $Z_{2}\times Z_{2}$ orientifold of the \mbox{Type 0B} theory, wich are similar to our case, have been studied in \cite{forger1}.\\
The model we are going to describe is without discrete torsion, Wilson lines and quantized $NS-NS$ antisymmetric tensor that would reduce the rank of the gauge groups.\\
We now set about breaking the supersymmetry with a Scherk-Schwarz deformation of the torus partition function.
In ordinary quantum field theory the SS mechanism consists in giving some fields a periodicity condition along a compactified dimension up to a symmetry transformation, $\Phi (x+2\pi R)=e^{2i\pi q\omega}\Phi(x)$, where $R$ is the radius of the dimension, $q$ the symmetry charge of the field and $\omega$ the parameter of the transformation \cite{ss1}.
The momenta of these fields in that dimension are then shifted proportionally to their symmetry charge, and their masses are accordingly shifted.
If the R-simmetry is used, this shift is generally different for the different components of the supermultiplets, yielding a spontaneous supersymmetry breaking.
The symmetry used in this paper is the spacetime fermion parity $(-1)^{F}$ along the tenth dimension \cite{antoniadis1}\cite{benakli1}.
The extension of this method to closed superstrings has been developed in \cite{ss2}, where it is shown how the shift in the momenta has to be accompanied by a modification of the compactification lattice in order to preserve modular invariance, yielding a deformation of the torus partition function.
The result in our case is that the fermionic part of (\ref{toron}), i.e.
\beq
\sum_{\ba{c} spin \\ structures \ea }C \l( \ba{c} \bf{a} \\ \bf{b} \ea \r) \wt{C} \l( \ba{c} \bf{a} \\ \bf{b} \ea \r) \th^{4} \l( \ba{c} \bf{a} \\ \bf{b} \ea \r) (\t) \ \wt{\th}^{4} \l( \ba{c} \bf{a} \\ \bf{b} \ea \r) (\wt{\t})~,
\eeq
where the coefficients $C$ are phases given by the GSO projection, is changed by the temperature-like Scherk-Schwarz deformation $(-1)^{F}$ in:
\bea \label{deformazione}
&& \sum_{ \ba{c} spin \\ structures \ea }  C^{*}\l( \ba{c} {\bf a}_{L} \\ {\bf b}_{L} \ea \r)\wt{C}^{*}\l( \ba{c} {\bf a}_{R} \\ {\bf b}_{R} \ea \r)\times \nonumber\\
&& \times \prod_{i_{L}=1}^{4}\th \l( \ba{c} {\bf a}_{i_{L}}- n {\bf \omega}_{i_{L}} \\ {\bf b}_{i_{L}}+ m {\bf \omega}_{i_{L}} \ea \r)(\t)\prod_{i_{R}=1}^{4}\wt{\th}\l( \ba{c} {\bf a}_{i_{R}}- n {\bf \omega}_{i_{R}} \\ {\bf b}_{i_{R}}+ m {\bf \omega}_{i_{R}} \ea \r)(\wt{\t})~,
\eea
with
\beq \label{deformazione1}
C^{*}\l( \ba{c} \bf{a} \\ \bf{b} \ea \r)=e^{2i\pi n{\bf \omega}({\bf b}+\frac{m}{2}{\bf \omega})}C\l( \ba{c} \bf{a} \\ \bf{b} \ea \r)~.
\eeq
In (\ref{deformazione}) and (\ref{deformazione1}), ${\bf \omega}_{i_{L,R}}$ are the parameters of the transformation (one for each $\theta $-function), in our case equal to ${\bf \omega}_{L}={\bf \omega}_{R}=(0,0,0,1)$.
We proceed now to the building of the open descendant of the Type-IIB \mbox{SS-deformed} theory following the procedure of \cite{sagnotti1}, according to  the standard methods \cite{sagnotti5}.\\
Introducing the level one O(2) characters:
\bea \label{eq:caratteri2}
I_{2}&=&\frac{\th_{3}+\th_{4}}{2 \eta}~,\qquad ~
V_{2}=\frac{\th_{3}-\th_{4}}{2 \eta}~,\nn \\
S_{2}&=&\frac{\th_{2}+i\th_{1}}{2 \eta}~,\qquad
C_{2}=\frac{\th_{2}-i\th_{1}}{2 \eta}~,
\eea
we can rewrite the deformed torus partition function in terms of the following combinations:
\bea \label{caratteria}
\tau_{oo}&=& V_{2}I_{2}I_{2}I_{2} + I_{2}V_{2}V_{2}V_{2} - S_{2}S_{2}S_{2}S_{2} - C_{2}C_{2}C_{2}C_{2}\,\nn\\
\tau_{og}&=& I_{2}V_{2}I_{2}I_{2} + V_{2}I_{2}V_{2}V_{2} - C_{2}C_{2}S_{2}S_{2} - S_{2}S_{2}C_{2}C_{2}\,\nn\\
\tau_{oh}&=& I_{2}I_{2}I_{2}V_{2} + V_{2}V_{2}V_{2}I_{2} - C_{2}S_{2}S_{2}C_{2} - S_{2}C_{2}C_{2}S_{2}\,\nn\\
\tau_{of}&=& I_{2}I_{2}V_{2}I_{2} + V_{2}V_{2}I_{2}V_{2} - C_{2}S_{2}C_{2}S_{2} - S_{2}C_{2}S_{2}C_{2}\,\nn\\
\tau_{go}&=& V_{2}I_{2}S_{2}C_{2} + I_{2}V_{2}C_{2}S_{2} - S_{2}S_{2}V_{2}I_{2} - C_{2}C_{2}I_{2}V_{2}\,\nn\\
\tau_{gg}&=& I_{2}V_{2}S_{2}C_{2} + V_{2}I_{2}C_{2}S_{2} - S_{2}S_{2}I_{2}V_{2} - C_{2}C_{2}V_{2}I_{2}\,\nn\\
\tau_{gh}&=& I_{2}I_{2}S_{2}S_{2} + V_{2}V_{2}C_{2}C_{2} - C_{2}S_{2}V_{2}V_{2} - S_{2}C_{2}I_{2}I_{2}\,\nn\\
\tau_{gf}&=& I_{2}I_{2}C_{2}C_{2} + V_{2}V_{2}S_{2}S_{2} - S_{2}C_{2}V_{2}V_{2} - C_{2}S_{2}I_{2}I_{2}\,\nn\\
\tau_{ho}&=& V_{2}S_{2}C_{2}I_{2} + I_{2}C_{2}S_{2}V_{2} - C_{2}I_{2}V_{2}C_{2} - S_{2}V_{2}I_{2}S_{2}\,\nn\\
\tau_{hg}&=& I_{2}C_{2}C_{2}I_{2} + V_{2}S_{2}S_{2}V_{2} - C_{2}I_{2}I_{2}S_{2} - S_{2}V_{2}V_{2}C_{2}\,\nn\\
\tau_{hh}&=& I_{2}S_{2}C_{2}V_{2} + V_{2}C_{2}S_{2}I_{2} - S_{2}I_{2}V_{2}S_{2} - C_{2}V_{2}I_{2}C_{2}\,\nn\\
\tau_{hf}&=& I_{2}S_{2}S_{2}I_{2} + V_{2}C_{2}C_{2}V_{2} - C_{2}V_{2}V_{2}S_{2} - S_{2}I_{2}I_{2}C_{2}\,\nn\\
\tau_{fo}&=& V_{2}S_{2}I_{2}C_{2} + I_{2}C_{2}V_{2}S_{2} - S_{2}V_{2}S_{2}I_{2} - C_{2}I_{2}C_{2}V_{2}\,\nn\\
\tau_{fg}&=& I_{2}C_{2}I_{2}C_{2} + V_{2}S_{2}V_{2}S_{2} - C_{2}I_{2}S_{2}I_{2} - S_{2}V_{2}C_{2}V_{2}\,\nn\\
\tau_{fh}&=& I_{2}S_{2}I_{2}S_{2} + V_{2}C_{2}V_{2}C_{2} - C_{2}V_{2}S_{2}V_{2} - S_{2}I_{2}C_{2}I_{2}\,\nn\\
\tau_{ff}&=& I_{2}S_{2}V_{2}C_{2} + V_{2}C_{2}I_{2}S_{2} - C_{2}V_{2}C_{2}I_{2} - S_{2}I_{2}S_{2}V_{2}\;
\eea
\bea \label{caratterib}
\tau^{'}_{oo}&=& V_{2}I_{2}I_{2}I_{2} + I_{2}V_{2}V_{2}V_{2} - C_{2}S_{2}S_{2}C_{2} - S_{2}C_{2}C_{2}S_{2}\,\nn\\
\tau^{'}_{og}&=& I_{2}V_{2}I_{2}I_{2} + V_{2}I_{2}V_{2}V_{2} - C_{2}S_{2}C_{2}S_{2} - S_{2}C_{2}S_{2}C_{2}\,\nn\\
\tau^{'}_{oh}&=& I_{2}I_{2}I_{2}V_{2} + V_{2}V_{2}V_{2}I_{2} - S_{2}S_{2}S_{2}S_{2} - C_{2}C_{2}C_{2}C_{2}\,\nn\\
\tau^{'}_{of}&=& I_{2}I_{2}V_{2}I_{2} + V_{2}V_{2}I_{2}V_{2} - C_{2}C_{2}S_{2}S_{2} - S_{2}S_{2}C_{2}C_{2}\,\nn\\
\tau^{'}_{go}&=& I_{2}I_{2}S_{2}C_{2} + V_{2}V_{2}C_{2}S_{2} - S_{2}S_{2}V_{2}V_{2} - C_{2}C_{2}I_{2}I_{2}\,\nn\\
\tau^{'}_{gg}&=& I_{2}I_{2}C_{2}S_{2} + V_{2}V_{2}S_{2}C_{2} - S_{2}S_{2}I_{2}I_{2} - C_{2}C_{2}V_{2}V_{2}\,\nn\\
\tau^{'}_{gh}&=& V_{2}I_{2}S_{2}S_{2} + I_{2}V_{2}C_{2}C_{2} - S_{2}C_{2}I_{2}V_{2} - C_{2}S_{2}V_{2}I_{2}\,\nn\\
\tau^{'}_{gf}&=& I_{2}V_{2}S_{2}S_{2} + V_{2}I_{2}C_{2}C_{2} - C_{2}S_{2}I_{2}V_{2} - S_{2}C_{2}V_{2}I_{2}\,\nn\\
\tau^{'}_{ho}&=& V_{2}S_{2}C_{2}I_{2} + I_{2}C_{2}S_{2}V_{2} - S_{2}I_{2}V_{2}S_{2} - C_{2}V_{2}I_{2}C_{2}\,\nn\\
\tau^{'}_{hg}&=& I_{2}C_{2}C_{2}I_{2} + V_{2}S_{2}S_{2}V_{2} - C_{2}V_{2}V_{2}S_{2} - S_{2}I_{2}I_{2}C_{2}\,\nn\\
\tau^{'}_{hh}&=& I_{2}S_{2}C_{2}V_{2} + V_{2}C_{2}S_{2}I_{2} - C_{2}I_{2}V_{2}C_{2} - S_{2}V_{2}I_{2}S_{2}\,\nn\\
\tau^{'}_{hf}&=& I_{2}S_{2}S_{2}I_{2} + V_{2}C_{2}C_{2}V_{2} - C_{2}I_{2}I_{2}S_{2} - S_{2}V_{2}V_{2}C_{2}\,\nn\\
\tau^{'}_{fo}&=& I_{2}S_{2}I_{2}C_{2} + V_{2}C_{2}V_{2}S_{2} - S_{2}V_{2}S_{2}V_{2} - C_{2}I_{2}C_{2}I_{2}\,\nn\\
\tau^{'}_{fg}&=& I_{2}S_{2}V_{2}S_{2} + V_{2}C_{2}I_{2}C_{2} - C_{2}I_{2}S_{2}V_{2} - S_{2}V_{2}C_{2}I_{2}\,\nn\\
\tau^{'}_{fh}&=& V_{2}S_{2}I_{2}S_{2} + I_{2}C_{2}V_{2}C_{2} - S_{2}I_{2}C_{2}V_{2} - C_{2}V_{2}S_{2}I_{2}\,\nn\\
\tau^{'}_{ff}&=& I_{2}C_{2}I_{2}S_{2} + V_{2}S_{2}V_{2}C_{2} - S_{2}I_{2}S_{2}I_{2} - C_{2}V_{2}C_{2}V_{2}\;
\eea
\bea \label{caratteric}
\sigma_{oo}&=& I_{2}I_{2}I_{2}I_{2} + V_{2}V_{2}V_{2}V_{2} - C_{2}S_{2}S_{2}S_{2} - S_{2}C_{2}C_{2}C_{2}\,\nn\\
\sigma_{og}&=& I_{2}I_{2}V_{2}V_{2} + V_{2}V_{2}I_{2}I_{2} - S_{2}C_{2}S_{2}S_{2} - C_{2}S_{2}C_{2}C_{2}\,\nn\\
\sigma_{oh}&=& I_{2}V_{2}V_{2}I_{2} + V_{2}I_{2}I_{2}V_{2} - S_{2}S_{2}S_{2}C_{2} - C_{2}C_{2}C_{2}S_{2}\,\nn\\
\sigma_{of}&=& I_{2}V_{2}I_{2}V_{2} + V_{2}I_{2}V_{2}I_{2} - C_{2}C_{2}S_{2}C_{2} - S_{2}S_{2}C_{2}S_{2}\,\nn\\
\sigma_{ho}&=& I_{2}S_{2}C_{2}I_{2} + V_{2}C_{2}S_{2}V_{2} - S_{2}I_{2}V_{2}C_{2} - C_{2}V_{2}I_{2}S_{2}\,\nn\\
\sigma_{hg}&=& I_{2}S_{2}S_{2}V_{2} + V_{2}C_{2}C_{2}I_{2} - S_{2}I_{2}I_{2}S_{2} - C_{2}V_{2}V_{2}C_{2}\,\nn\\
\sigma_{hh}&=& I_{2}C_{2}S_{2}I_{2} + V_{2}S_{2}C_{2}V_{2} - S_{2}V_{2}I_{2}C_{2} - C_{2}I_{2}V_{2}S_{2}\,\nn\\
\sigma_{hf}&=& I_{2}C_{2}C_{2}V_{2} + V_{2}S_{2}S_{2}I_{2} - C_{2}I_{2}I_{2}C_{2} - S_{2}V_{2}V_{2}S_{2}~.
\eea
We will use the notation ($i=o,g,h,f$):
\bea \label{iT}
T_{io}&=& \tau_{io}+\tau_{ig}+\tau_{ih}+\tau_{if}~,\qquad T_{ig}= \tau_{io}+\tau_{ig}-\tau_{ih}-\tau_{if}~,\nn \\
T_{ih}&=& \tau_{io}-\tau_{ig}+\tau_{ih}-\tau_{if}~,\qquad T_{if}= \tau_{io}-\tau_{ig}-\tau_{ih}+\tau_{if}~,
\eea
and likewise for the $\tau^{'}$'s ($T^{'}_{ij}$) and the $\sigma$'s ($S_{ij}$); moreover a superscript ``F'' or ``B'' for the T's will denote the Fermionic or Bosonic part of the characters. 
We will also need the definitions
\bea
E_{0}&=&\sum_{m,n}\frac{1+(-1)^{n}}{2}Z_{m,n}~, \qquad \quad O_{0}=\sum_{m,n}\frac{1-(-1)^{n}}{2}Z_{m,n}~,\nn\\
E_{\frac{1}{2}}&=&\sum_{m,n}\frac{1+(-1)^{n}}{2}Z_{m+\frac{1}{2},n}~, \qquad O_{\frac{1}{2}}=\sum_{m,n}\frac{1-(-1)^{n}}{2}Z_{m+\frac{1}{2},n}~,
\eea
where
\beq
Z_{m+a,n}=\frac{1}{|\eta(\t)|^{2}}\sum_{m,n}q^{\frac{1}{2}(\frac{m+a}{R}+\frac{nR}{2})^{2}}\wt{q}^{\frac{1}{2}(\frac{m+a}{R}-\frac{nR}{2})^{2}}~,
\eeq
and
\beq
Z_{m+a}=\frac{1}{\eta}\sum_{m}q^{\frac{1}{2}(\frac{m+a}{R})^{2}}~, \qquad
\wt{Z}_{n+b}=\frac{1}{\eta}\sum_{n}q^{\frac{1}{2}(\frac{(n+b)R}{2})^{2}}~.
\eeq
In the whole paper $\a ^{'}=2$, so that $Z_{2m}$ and $Z_{2m+1}$ denote the sums over integer and half-integer momenta, and likewise for the winding numbers.\\
The torus partition function results:
\bea \label{toro}
T &=& \frac{1}{4}\l\{\Lambda_{1}\Lambda_{2}Z_{m,n}\l[ E_{0}(|T^{B}_{oo}|^{2}+|T^{F}_{oo}|^{2})+O_{0}(|S^{B}_{oo}|^{2}+|S^{F}_{oo}|^{2})+E_{\frac{1}{2}}(T^{B}_{oo}\wt{T}^{F}_{oo}+\wt{T}^{B}_{oo}T^{F}_{oo}) \r. \r. \nn \\
&& \l. \l. +O_{\frac{1}{2}}(S^{B}_{oo}\wt{S}^{F}_{oo}+\wt{S}^{B}_{oo}S^{F}_{oo}) \r]
+\Lambda_{1}|\frac{4\eta^{2}}{\th^{2}_{2}}|^{2}\frac{1}{2}\l[|T_{og}|^{2}+|T^{'}_{og}|^{2}\r] + \Lambda_{2}|\frac{4\eta^{2}}{\th^{2}_{2}}|^{2}\frac{1}{2}\l[|T_{of}|^{2}+|T^{'}_{of}|^{2}\r]  \r. \nn\\
&& \l. +|\frac{4\eta^{2}}{\th^{2}_{2}}|^{2}Z_{m,n}\l[ E_{0}(|T^{B}_{oh}|^{2}+|T^{F}_{oh}|^{2})+O_{0}(|S^{B}_{oh}|^{2}+|S^{F}_{oh}|^{2})+E_{\frac{1}{2}}(T^{B}_{oh}\wt{T}^{F}_{oh}+\wt{T}^{B}_{oh}T^{F}_{oh}) \r. \r. \nn\\
&& \l. \l. +O_{\frac{1}{2}}(S^{B}_{oh}\wt{S}^{F}_{oh}+\wt{S}^{B}_{oh}S^{F}_{oh})\r] +
\Lambda_{1}|\frac{4\eta^{2}}{\th^{2}_{4}}|^{2}\frac{1}{2}\l[|T_{go}|^{2}+|T^{'}_{go}|^{2}\r] + \Lambda_{2}|\frac{4\eta^{2}}{\th^{2}_{4}}|^{2}\frac{1}{2}\l[|T_{fo}|^{2}+|T^{'}_{fo}|^{2}\r]\r. \nn\\
&& \l. + |\frac{4\eta^{2}}{\th^{2}_{4}}|^{2}Z_{m,n}\l[ E_{0}(|T^{B}_{ho}|^{2}+|T^{F}_{ho}|^{2})+O_{0}(|S^{B}_{ho}|^{2}+|S^{F}_{ho}|^{2})+E_{\frac{1}{2}}(T^{B}_{ho}\wt{T}^{F}_{ho}+\wt{T}^{B}_{ho}T^{F}_{ho}) \r. \r. \nn\\
&& \l. \l. +O_{\frac{1}{2}}(S^{B}_{ho}\wt{S}^{F}_{ho}+\wt{S}^{B}_{ho}S^{F}_{ho})\r] +
\Lambda_{1}|\frac{4\eta^{2}}{\th^{2}_{3}}|^{2}\frac{1}{2}\l[|T_{gg}|^{2}+|T^{'}_{gg}|^{2}\r] + \Lambda_{2}|\frac{4\eta^{2}}{\th^{2}_{3}}|^{2}\frac{1}{2}\l[|T_{ff}|^{2}+|T^{'}_{ff}|^{2}\r] \r. \nn\\
&& \l. + |\frac{4\eta^{2}}{\th^{2}_{3}}|^{2}Z_{m,n}\l[ E_{0}(|T^{B}_{hh}|^{2}+|T^{F}_{hh}|^{2})+O_{0}(|S^{B}_{hh}|^{2}+|S^{F}_{hh}|^{2})-E_{\frac{1}{2}}(T^{B}_{hh}\wt{T}^{F}_{hh}+\wt{T}^{B}_{hh}T^{F}_{hh}) \r. \r. \nn\\
&& \l. \l. +O_{\frac{1}{2}}(S^{B}_{hh}\wt{S}^{F}_{hh}+\wt{S}^{B}_{hh}S^{F}_{hh})\r] +
|\frac{8\eta^{3}}{\th_{2}\th_{3}\th_{4}}|^{2}\frac{1}{2}\l[|T_{gh}|^{2}+|T^{'}_{gh}|^{2}+|T_{gf}|^{2}+|T^{'}_{gf}|^{2}+|T_{hg}|^{2}\r. \r. \nn\\
&& \l. \l. +|T^{'}_{hg}|^{2}+|T_{hf}|^{2}+|T^{'}_{hf}|^{2}+|T_{fg}|^{2}+|T^{'}_{fg}|^{2}+|T_{fh}|^{2}+|T^{'}_{fh}|^{2} \r]  \r\}~,
\eea
where we left implicit the contribution of the transverse bosons and the argument of the characters, $q=exp(2i\pi\tau)$, with $\tau$ the modulus of the torus.
The same will be done for the other amplitudes, namely the Klein bottle, annulus and M\"{o}bius strip.
In these cases the relations between direct and transverse channel moduli are the ones in \cite{sagnotti2}.\\
The direct channel Klein bottle amplitude reads:
\bea \label{klein}
K &=& \frac{1}{8}\l\{ P_{1}P_{2}P_{3}T_{oo} + P_{1}W_{2}\wt{Z}_{n}(\wt{Z}_{2n}T_{oo}+\wt{Z}_{2n+1}S_{oo}) + W_{1}P_{2}\wt{Z}_{n}(\wt{Z}_{2n}T_{oo}+\wt{Z}_{2n+1}S_{oo})\r. \nn\\
&& \l.  + W_{1}W_{2}P_{3}T_{oo} +
16\l(\frac{\eta}{\th_{4}}\r)^{2} \l[ (P_{1}+W_{1})(T_{go}+T^{'}_{go}) + (P_{2}+W_{2})(T_{fo}+T^{'}_{fo})\r. \r. \nn\\
&& \l. \l. + 2P_{3}T_{ho} + 2\wt{Z}_{n}(\wt{Z}_{2n}T_{ho}+\wt{Z}_{2n+1}S_{ho}) \r] \r\}~.
\eea
$P_{i}$ ($W_{i}$) denotes the restriction of $\Lambda_{i}$ to its momentum (winding) sublattice as in \cite{sagnotti2}\cite{sagnotti3}\cite{sagnotti4}.\\
A modular S-transformation \cite{sagnotti1} gives the transverse channel Klein bottle amplitude:
\bea \label{kleint}
\wt{K} &=& \frac{2^{5}}{8}\l\{ v_{1}v_{2}v_{3}W^{e}_{1}W^{e}_{2}W^{e}_{3}T_{oo} + \frac{v_{1}}{v_{2}v_{3}}W^{e}_{1}P^{e}_{2}Z_{2m}(Z_{2m}T^{B}_{oo} +Z_{2m+1}T^{F}_{oo})\r. \nn\\
&& \l.  +  \frac{v_{2}}{v_{1}v_{3}}P^{e}_{1}W^{e}_{2}Z_{2m}(Z_{2m}T^{B}_{oo}+Z_{2m+1}T^{F}_{oo}) + \frac{v_{3}}{v_{1}v_{2}}P^{e}_{1}P^{e}_{2}W^{e}_{3}T_{oo}\r. \nn\\
&& \l. +\l(\frac{2\eta}{\th_{2}}\r)^{2} \l[ (v_{1}W^{e}_{1}+\frac{P^{e}_{1}}{v_{1}})(T_{og}+T^{'}_{og}) +  (v_{2}W^{e}_{2}+\frac{P^{e}_{2}}{v_{2}})(T_{of}+T^{'}_{of}) + 2v_{3}W^{e}_{3}T_{oh}\r. \r. \nn\\
&& \l. \l. + 2\frac{Z_{2m}}{v_{3}}(Z_{2m}T^{B}_{oh}+Z_{2m+1}T^{F}_{oh}) \r] \r\}~,
\eea
where the $v_{i}$ denote the volumes of the compactified tori.
The contributions at the origin of the lattice can be organized in terms whose coefficients are perfect squares.
Furthermore, there is no need to introduce D$\wt{5}$-branes as in \cite{sagnotti4} (and \cite{sagnotti1}) since in this case there is no change in the chirality of the fermions in the deformed theory.\\
The only characters that can descend on the transverse channel annulus amplitude are the $T_{oi}$, $T^{'}_{oi}$ and $S_{oi}$, since the others are not self-conjugate.
The presence of two types of characters ($T_{oi}$, $T^{'}_{oi}$) produces the separation of two of the three sets of D5-branes into subsets, that we will call $D_{1a}$, $D_{1b}$, $D_{2a}$ and $D_{2b}$.
If we choose a configuration of branes, all placed at fixed points, in which the third set ($D_{3}$) intersects all the subsets while the $D_{ia}$'s intersect the $D_{ib}$'s, we have:
\bea \label{anellot}
\wt{A} &=& \frac{2^{-5}}{8}\l\{ N^{2}v_{1}v_{2}v_{3}W_{1}W_{2}\wt{Z}_{n}(\wt{Z}_{2n}T_{oo}+\wt{Z}_{2n+1}S_{oo})\r. \nn\\
&& \l. + \frac{v_{1}}{v_{2}v_{3}}W_{1}P_{2}Z_{m}\l[(D_{1a}+D_{1b})^{2}(Z_{2m}T^{B}_{oo}+Z_{2m+1}T^{F}_{oo})\r. \r. \nn\\
&& \l. \l. +(D_{1a}-D_{1b})^{2}(Z_{2m+1}T^{B}_{oo}+Z_{2m}T^{F}_{oo})\r] \r. \nn\\
&& \l. +\frac{v_{2}}{v_{1}v_{3}}P_{1}W_{2}Z_{m}\l[(D_{2a}+D_{2b})^{2}(Z_{2m}T^{B}_{oo}+Z_{2m+1}T^{F}_{oo})\r. \r. \nn\\
&& \l. \l. +(D_{2a}-D_{2b})^{2}(Z_{2m+1}T^{B}_{oo}+Z_{2m}T^{F}_{oo}) \r] + D^{2}_{3}\frac{v_{3}}{v_{1}v_{2}}P_{1}P_{2}\wt{Z}_{n}(\wt{Z}_{2n}T_{oo}+\wt{Z}_{2n+1}S_{oo})\r. \nn\\
&& \l. +\l(\frac{2\eta}{\th_{2}}\r)^{2} \l[ v_{1}W_{1}(2ND_{1a}T_{og}+2ND_{1b}T^{'}_{og}) +  v_{2}W_{2}(2ND_{2a}T_{of}+2ND_{2b}T^{'}_{of})\r. \r. \nn\\
&& \l. \l. + v_{3}\wt{Z}_{n}2ND_{3}(\wt{Z}_{2n}T_{oh}+\wt{Z}_{2n+1}S_{oh})  \r] + 
\l(\frac{2\eta}{\th_{2}}\r)^{2} \l[ \frac{1}{v_{1}}P_{1}(2D_{2a}D_{3}T_{og}+2D_{2b}D_{3}T^{'}_{og})\r. \r. \nn\\
&& \l. \l. +\frac{1}{v_{2}}P_{2}(2D_{1a}D_{3}T_{of}+2D_{1b}D_{3}T^{'}_{of}) \r. \r. \nn\\
&& \l. \l. + \frac{1}{v_{3}}Z_{m}\l( (2D_{1a}D_{2a}+2D_{1a}D_{2b}+2D_{2a}D_{2b}+2D_{2a}D_{1b})(Z_{2m}T^{B}_{oh}+Z_{2m+1}T^{F}_{oh}) \r. \r. \r. \nn\\
&& \l. \l. \l. +(2D_{1a}D_{2a}-2D_{1a}D_{2b}+2D_{2a}D_{2b}-2D_{2a}D_{1b})(Z_{2m+1}T^{B}_{oh}+Z_{2m}T^{F}_{oh}) \r) \r] \r\}~.
\eea
Also this amplitude can be organized as sums of perfect squares.
The direct channel can be obtained by an S-transformation, and reads:
\bea \label{anello}
A &=& \frac{1}{8}\l\{ N^{2}P_{1}P_{2}Z_{m}(Z_{2m}T^{B}_{oo}+Z_{2m+1}T^{F}_{oo}) \r. \nn\\
&& \l. + P_{1}W_{2}\wt{Z}_{n}\l[(D_{1a}^{2}+D_{1b}^{2})\wt{Z}_{2n}T_{oo}+2D_{1a}D_{1b}\wt{Z}_{2n+1}S_{oo})\r]\r. \nn\\
&& \l. + W_{1}P_{2}\wt{Z}_{n}\l[(D_{2a}^{2}+D_{2b}^{2})\wt{Z}_{2n}T_{oo}+2D_{2a}D_{2b}\wt{Z}_{2n+1}S_{oo})\r] \r. \nn\\
&& \l. + D_{3}^{2}W_{1}W_{2}Z_{m}(Z_{2m}T^{B}_{oo}+Z_{2m+1}T^{F}_{oo}) +
\l(\frac{\eta}{\th_{4}}\r)^{2} \l[ P_{1}(2ND_{1a}T_{go}+2ND_{1b}T^{'}_{go}) \r. \r. \nn\\
&& \l. \l. + P_{2}(2ND_{2a}T_{fo}+2ND_{2b}T^{'}_{fo}) + Z_{m}2ND_{3}(Z_{2m}T^{B}_{ho}+Z_{2m+1}T^{F}_{ho})  \r] \r. \nn\\
&& \l. + \l(\frac{\eta}{\th_{4}}\r)^{2} \l[ W_{1}(2D_{2a}D_{3}T_{go}+2D_{2b}D_{3}T^{'}_{go}) + W_{2}(2D_{1a}D_{3}T_{fo}+2D_{1b}D_{3}T^{'}_{fo}) \r. \r. \nn\\
&& \l. \l. + \wt{Z}_{n}\l( (2D_{1a}D_{1b}+2D_{2a}D_{2b})\wt{Z}_{2n}T_{ho}+(2D_{1a}D_{2b}+2D_{2a}D_{1b})\wt{Z}_{2n+1}S_{ho} \r) \r]  
\r\}~.
\eea
From the geometric mean of the amplitudes $\wt{K}$ and $\wt{A}$ we can derive the transverse channel M\"obius amplitude:
\bea \label{moebiust}
\wt{M} &=& -\frac{1}{4}\l\{ Nv_{1}v_{2}v_{3}W^{e}_{1}W^{e}_{2}\wt{Z}_{2n}\l[(\wt{Z}_{4n}+\wt{Z}_{4n+2})\hat{T}^{B}_{oo}+(\wt{Z}_{4n}-\wt{Z}_{4n+2})\hat{T}^{F}_{oo}\r] \r. \nn\\
&& \l.+ \frac{v_{1}}{v_{2}v_{3}}W^{e}_{1}P^{e}_{2}Z_{2m}(D_{1a}+D_{1b})(Z_{2m}\hat{T}^{B}_{oo}+Z_{2m+1}\hat{T}^{F}_{oo}) \r. \nn\\
&& \l. + \frac{v_{2}}{v_{1}v_{3}}P^{e}_{1}W^{e}_{2}Z_{2m}(D_{2a}+D_{2b})(Z_{2m}\hat{T}^{B}_{oo}+Z_{2m+1}\hat{T}^{F}_{oo}) \r. \nn\\
&& \l. + D_{3}\frac{v_{3}}{v_{1}v_{2}}P^{e}_{1}P^{e}_{2}\wt{Z}_{2n}\l[(\wt{Z}_{4n}+\wt{Z}_{4n+2})\hat{T}^{B}_{oo}+(\wt{Z}_{4n}-\wt{Z}_{4n+2})\hat{T}^{F}_{oo} \r] \r. \nn\\
&& \l. +\l(\frac{2\hat{\eta}}{\hat{\th}_{2}}\r)^{2} \l[ v_{1}W^{e}_{1}(\frac{N}{2}\hat{T}_{og}+\frac{N}{2}\hat{T}^{'}_{og}+D_{1a}\hat{T}_{og}+D_{1b}\hat{T}^{'}_{og})\r. \r. \nn\\
&& \l. \l. +v_{2}W^{e}_{2}(\frac{N}{2}\hat{T}_{of}+\frac{N}{2}\hat{T}^{'}_{of}+D_{2a}\hat{T}_{of}+D_{2b}\hat{T}^{'}_{of}) \r. \r. \nn\\
&& \l. \l. + v_{3}\wt{Z}_{2n}(N+D_{3})\l( (\wt{Z}_{4n}+\wt{Z}_{4n+2})\hat{T}^{B}_{oh}+(\wt{Z}_{4n}-\wt{Z}_{4n+2})\hat{T}^{F}_{oh}\r)  \r] \r. \nn\\
&& \l. +\l(\frac{2\hat{\eta}}{\hat{\th}_{2}}\r)^{2} \l[ \frac{1}{v_{1}}P^{e}_{1}(\frac{D_{3}}{2}\hat{T}_{og}+\frac{D_{3}}{2}\hat{T}^{'}_{og}+D_{2a}\hat{T}_{og}+D_{2b}\hat{T}^{'}_{og})\r. \r. \nn\\
&& \l. \l. +\frac{1}{v_{2}}P^{e}_{2}(\frac{D_{3}}{2}\hat{T}_{of}+\frac{D_{3}}{2}\hat{T}^{'}_{of}+D_{1a}\hat{T}_{of}+D_{1b}\hat{T}^{'}_{of})\r. \r. \nn\\
&& \l. \l. + \frac{1}{v_{3}}Z_{2m} \l( (D_{1a}+D_{1b}+D_{2a}+D_{2b})Z_{2m}\hat{T}^{B}_{oh} \r. \r. \r. \nn\\
&& \l. \l. \l. +(D_{1a}+D_{1b}-D_{2a}-D_{2b})Z_{2m+1}\hat{T}^{F}_{oh} \r)   \r]  
\r\}~,
\eea
where we have used a proper basis of ``hatted characters'' \cite{sagnotti3}, while the ``hats'' on the $Z$'s, $\wt{Z}$'s, $P$'s and $W$'s have been left implicit.
In order to obtain the direct channel contribution we have to make a P-transformation \cite{sagnotti1}, that gives:
\bea \label{moebius}
M &=& -\frac{1}{8}\l\{ NP_{1}P_{2}Z_{m}(Z_{2m}\hat{T}^{B}_{oo}+Z_{2m+1}\hat{T}^{F}_{oo}) \r. \nn\\
&& \l. + P_{1}W_{2}\wt{Z}_{n}(D_{1a}+D_{1b})(\wt{Z}_{2n}\hat{T}^{B}_{oo}+(-1)^{n}\wt{Z}_{2n}\hat{T}^{F}_{oo}) \r. \nn\\
&& \l. + W_{1}P_{2}\wt{Z}_{n}(D_{2a}+D_{2b})(\wt{Z}_{2n}\hat{T}^{B}_{oo}+(-1)^{n}\wt{Z}_{2n}\hat{T}^{F}_{oo}) + D_{3}W_{1}W_{2}Z_{m}(Z_{2m}\hat{T}^{B}_{oo}+Z_{2m+1}\hat{T}^{F}_{oo})\r. \nn\\
&& \l. -\l(\frac{2\hat{\eta}}{\hat{\th}_{2}}\r)^{2} \l[ P_{1}(\frac{N}{2}\hat{T}_{og}+\frac{N}{2}\hat{T}^{'}_{og}+D_{1a}\hat{T}_{og}+D_{1b}\hat{T}^{'}_{og}) \r.  \r. \nn\\
&& \l. \l.  +P_{2}(\frac{N}{2}\hat{T}_{of}+\frac{N}{2}\hat{T}^{'}_{of}+D_{2a}\hat{T}_{of}+D_{2b}\hat{T}^{'}_{of})  + Z_{m}(N+D_{3})(Z_{2m}\hat{T}^{B}_{oh}+Z_{2m+1}\hat{T}^{F}_{oh})  \r] \r. \nn\\ 
&& \l. -\l(\frac{2\hat{\eta}}{\hat{\th}_{2}}\r)^{2} \l[ W_{1}(\frac{D_{3}}{2}\hat{T}_{og}+\frac{D_{3}}{2}\hat{T}^{'}_{og}+D_{2a}\hat{T}_{og}+D_{2b}\hat{T}^{'}_{og})\r. \r. \nn\\
&&  \l. \l. + W_{2}(\frac{D_{3}}{2}\hat{T}_{of}+\frac{D_{3}}{2}\hat{T}^{'}_{of}+D_{1a}\hat{T}_{of}+D_{1b}\hat{T}^{'}_{of}) + \wt{Z}_{n} \l( (D_{1a}+D_{1b}+D_{2a}+D_{2b})\wt{Z}_{2n}\hat{T}^{B}_{oh} \r. \r. \r. \nn\\
&& \l. \l. \l. +(D_{1a}+D_{1b}-D_{2a}-D_{2b})(-1)^{n}\wt{Z}_{2m}\hat{T}^{F}_{oh} \r)   \r] \r\}~.
\eea
Tadpole cancellation conditions give the results $N=D_{3}=32$ and $D_{1a}=D_{1b}=D_{2a}=D_{2b}=16$, so that in the model there are 32 D9-branes and three sets of 32 D5-branes each as in the supersymmetric case, but two of the groups of D5-branes are separated in two subsets of 16 branes each.
The model is non chiral, therefore free of gauge and \mbox{gravitational} anomalies.
Inspection of the M\"obius amplitude reveals the need for symplectic groups, and the correct Chan-Paton parametrization $N=2n$ and $D_{i}=2d_{i}$ gives the symmetry groups $Usp(16)_{9}\times Usp(8)_{5_{1}}\times Usp(8)_{5_{1}}\times Usp(8)_{5_{2}}\times Usp(8)_{5_{2}}\times Usp(16)_{5_{3}}$.
The massless spectrum contains only the bosonic part of the $N=1$ adjoint vector multiplets and of the three chiral multiplets in the ({\bf 120}) for the groups $Usp(16)_{9}$ and $Usp(16)_{5_{3}}$, while the correspondent fermions are in this case massive.
For the remaining groups we have instead the full adjoint vector multiplets and chiral multiplets in the ({\bf 28},{\bf 1}) and ({\bf 1},{\bf 28}) of each $Usp(8)\times Usp(8)$.
We also have additional chiral multiplets in the ({\bf 16},{\bf 8}) for the sectors $ND_{1}$, $ND_{2}$, $D_{3}D_{1}$, $D_{3}D_{2}$ (two each), in the ({\bf 8},{\bf 8}) for the sectors $D_{1}D_{2}$ (two), and only the scalars of the chiral multiplet in the ({\bf 16},{\bf 16}) for the sector $ND_{3}$.
The spectrum is therefore non supersymmetric in the sectors related to D-branes tangent to the breaking dimension, while it has the supersymmetry preserved on the other branes and in the mixed sectors, where nevertheless the symmetry groups are broken in two equal parts.

\medskip
\section*{Acknowledgments.}

\noindent 
I would like to thank Prof. Luciano Girardello for having introduced me to the study of string theory. 

\end{document}